\documentclass[12pt]{article}
\usepackage{latexsym}
\usepackage{epsfig,amssymb,euscript}
\usepackage{amsmath}
\numberwithin{equation}{section}  

\newsavebox{\ns}
\newsavebox{\dbrane}
\newsavebox{\dbshort}

\def\be{\begin{equation}}
\def\ee{\end{equation}}
\def\bea{\begin{eqnarray}}
\def\eea{\end{eqnarray}}

\def\Dslash{\,\,{\raise.15ex\hbox{/}\mkern-12mu D}}
\def\Dbarslash{\,\,{\raise.15ex\hbox{/}\mkern-12mu {\bar D}}}
\def\delslash{\,\,{\raise.15ex\hbox{/}\mkern-9mu \partial}}
\def\delbarslash{\,\,{\raise.15ex\hbox{/}\mkern-9mu {\bar\partial}}}
\def\pslash{\,\,{\raise.15ex\hbox{/}\mkern-9mu p}}
\def\calDslash{\,\,{\raise.15ex\hbox{/}\mkern-12mu {\cal D}}}

\newcommand\R{\mathbb{R}}
\newcommand\Z{\mathbb{Z}}

\newcommand\C{\mathbb{C}}

\newcommand\diff{\mathrm{d}}
\newcommand{\de}{\partial}

\newcommand{\vol}{\mathrm{vol}}

\begin{document}
\begin{titlepage}
\begin{center}
\today
{\small\hfill hep-th/0607080}\\
{\small\hfill Imperial/TP/2006/JG/01}\\
{\small\hfill CERN-PH-TH/2006-129}\\
{\small\hfill HUTP-06/A0018}\\

\vskip 2.5 cm 
{\Large \bf Obstructions to the Existence of}\\
\vskip 4mm
{\Large \bf Sasaki--Einstein Metrics}

\vskip 7mm
{Jerome P. Gauntlett$^{1,2}$, Dario Martelli$^{3}$, James Sparks$^{4,5}$ and Shing-Tung Yau$^{4}$}\\
\vskip 1 cm

1: Blackett Laboratory, Imperial College\\
London SW7 2AZ, U.K.\\
\vskip 0.5cm
2: The Institute for Mathematical Sciences, Imperial College\\
London SW7 2PG, U.K.\\
\vskip 0.5cm
3: Department of Physics, CERN Theory Unit\\
1211 Geneva 23, Switzerland\\
\vskip 0.5cm
4: Department of Mathematics, Harvard University \\
One Oxford Street, Cambridge, MA 02138, U.S.A.\\
\vskip 0.5cm
5: Jefferson Physical Laboratory, Harvard University \\
Cambridge, MA 02138, U.S.A.\\

\vskip 0.5cm

\end{center}

\begin{abstract}
\noindent
We describe two simple obstructions to the existence 
of Ricci--flat K\"ahler cone metrics on isolated Gorenstein 
singularities or, equivalently, to the existence of 
Sasaki--Einstein metrics
on the links of these singularities.
In particular, this also leads to new obstructions for K\"ahler--Einstein 
metrics on Fano orbifolds. We present several  families 
of hypersurface singularities that are obstructed, including 
 3--fold and 4--fold singularities of ADE type that have been 
studied previously in the physics literature. We show that 
the AdS/CFT dual of one obstruction is that the R--charge of 
a gauge invariant chiral primary operator violates the unitarity bound.

\end{abstract}

\end{titlepage}
\pagestyle{plain}
\setcounter{page}{1}
\newcounter{bean}
\baselineskip18pt

\tableofcontents


\section{Introduction}

The study of string theory and M--theory on singular manifolds is a very
rich subject that has led to many important insights. For  
geometries that develop an isolated singularity, one can 
model the local behaviour using a non--compact manifold. 
In this case, a natural geometric boundary condition is for the 
metric to asymptote to a cone away from the singularity. This means 
that one studies a family of metrics that asymptotically 
approach the conical form
\be\label{cone}
g_X = \diff r^2 + r^2 g_L\ee
with $(L,g_L)$ a compact Riemannian 
manifold. The dynamics of string theory or M--theory on 
special holonomy manifolds that are developing an isolated conical singularity 
$X$, with metric (\ref{cone}), has proved to be an extremely intricate subject. 

A particularly interesting setting is in the 
context of the AdS/CFT correspondence 
\cite{Maldacena}. 
The worldvolume theory of a 
large number of D3--branes placed at an isolated conical Calabi--Yau 3--fold 
singularity is expected to flow, at low energies, 
to a four--dimensional $\mathcal{N}=1$ superconformal field theory. In this case, 
the AdS/CFT conjecture states that this theory is dual to type IIB 
string theory on 
AdS$_5\times L$ \cite{Kehagias, KW, acharya, MP}. Similar remarks apply to M--theory on 
conical eight--dimensional singularities with special holonomy, 
which lead to superconformal theories in three dimensions that are dual to 
AdS$_4\times L$, although far less is known about this situation.

The focus of this paper will be on conical Calabi--Yau singularities, 
by which we mean
Ricci--flat K\"ahler metrics of the conical form (\ref{cone}). This gives, 
by definition, a Sasaki--Einstein metric on the base
of the cone $L$. 
We tacitly assume that $L$ is simply--connected which, although not entirely
necessary, always ensures the existence of a globally defined 
Killing spinor on $L$. 
A central role is played by the Reeb vector field
\be
\xi = J\left(r\frac{\partial}{\partial r}\right)\ee
where $J$ denotes the complex structure tensor on the cone $X$. $\xi$ 
is holomorphic, Killing, and has constant norm on the link 
$L=\{r=1\}$ of the singularity at $r=0$. If the orbits 
of $\xi$ all close then $L$ has a $U(1)$ isometry, which 
necessarily acts locally freely, and the Sasakian structure is said to be 
either regular or quasi--regular if this action is free or not, respectively. 
The orbit space is in general a 
positively curved K\"ahler--Einstein orbifold $(V,g_V)$, which is a smooth 
manifold in the regular case. More generally, 
the generic orbits of $\xi$ need not close, in which case 
the Sasakian structure is said to be irregular. The AdS/CFT correspondence
maps the symmetry generated by the Reeb vector field to the 
R--symmetry of the dual CFT. Thus for the quasi--regular case the CFT has
a $U(1)$ R--symmetry, whereas for the irregular case it has a non--compact
$\R$ R--symmetry.

Given a Sasaki--Einstein manifold $(L,g_L)$, the cone 
$X$, as a complex variety, is an isolated Gorenstein singularity. 
If $X_0$ denotes $X$ with the singular point removed, we 
have $X_0=\R_+\times L$ with $r>0$ a coordinate on $\R_+$. 
$X$ being Gorenstein means simply that there exists 
a nowhere zero holomorphic 
$(n,0)$--form $\Omega$ on $X_0$. 
One may then turn things around and ask which isolated Gorenstein 
singularities admit Sasaki--Einstein metrics on their links. 
This is a question in algebraic geometry, and it is an extremely 
difficult one. 
To give some idea of how difficult this question is, 
let us focus on the quasi--regular case. Thus, suppose that $X$ has a holomorphic $\C^*$ action, with orbit 
space being a Fano\footnote{We define a Fano orbifold $V$ to be a 
compact K\"ahler orbifold, such that the cohomology 
class of the Ricci--form in $H^2(V;\R)$ 
is represented by a positive $(1,1)$--form on $V$.
} manifold, or Fano orbifold, $V$. 
Then existence of a Ricci--flat K\"ahler cone metric on $X$, 
with conical symmetry generated by $\R_+\subset \C^*$, is 
well known to be equivalent to finding a K\"ahler--Einstein 
metric on $V$ -- for a review, see \cite{BGreview}. Existence of K\"ahler--Einstein metrics on 
Fanos is a very subtle problem that is still unsolved. That is, 
a set of necessary and sufficient algebraic conditions on $V$ are not 
known in general. There are two well--known holomorphic 
obstructions, due to Matsushima \cite{Matsushima} and Futaki \cite{Futaki}. 
The latter was related to Sasakian geometry in \cite{MSY2} 
and is not in fact an
obstruction from the Sasaki--Einstein point view. Specifically, it is possible
to have a Fano $V$ that has non--zero Futaki invariant and thus 
does not admit a
K\"ahler-Einstein metric, but nevertheless the link of the total space of
the canonical bundle over $V$ can admit a Sasaki--Einstein metric -- the point
is simply that the Reeb vector field is not\footnote{This 
happens, for example, when $V=\mathbb{F}_1$ -- the first del Pezzo 
surface. In this case both the Matsushima and Futaki theorems obstruct 
existence of a K\"ahler--Einstein metric on $V$, 
but there is nevertheless an irregular Sasaki--Einstein metric on the link in 
the total space of the canonical bundle over $V$ \cite{MS}.} the one associated with the
canonical bundle over $V$.
It is also known that vanishing of 
these two obstructions is, in general,  
insufficient for there to exist a K\"ahler--Einstein 
metric on $V$. It has been conjectured in \cite{yau} that 
$V$ admits a K\"ahler--Einstein metric if and only if it is 
\emph{stable}; proving this conjecture is currently a major 
research programme in geometry -- see, for example, \cite{donaldson}.
Thus, one also 
expects the existence of Ricci--flat K\"ahler cone metrics on 
an isolated Gorenstein singularity $X$ to be a subtle problem. 
This issue has been overlooked in some of the
physics literature, and  it has sometimes been
incorrectly assumed, or stated, that such conical Calabi--Yau metrics
exist on particular singularities, as we shall
discuss later.

The Reeb vector field contains a significant amount of 
information about the 
metric. For a fixed $X_0=\R_+\times L$, 
the Reeb vector field $\xi$ for a Sasaki--Einstein metric on $L$ 
satisfies a variational problem that depends only on
the complex structure of $X$ \cite{MSY, MSY2}. 
This is the geometric analogue of $a$--maximisation \cite{IW} in four dimensional 
superconformal field theories.
This allows one, in principle and often in practice, to obtain $\xi$, 
and hence in particular the volume, 
of a Sasaki--Einstein metric on $L$ -- 
\emph{assuming that this metric exists}. 

Now, for any $(2n-1)$--dimensional Einstein manifold $(L,g_L)$ with $\mathrm{Ric}=2(n-1)g_L$,
Bishop's theorem \cite{bishop} (see also \cite{Besse}) implies that the volume of $L$ 
is bounded from above by that of the round unit radius sphere. 
Thus we are immediately led to what we will call the {\bf Bishop obstruction} 
to the existence of Sasaki--Einstein metrics:
\begin{quote}
\begin{itshape}
  If the volume of the putative Sasaki--Einstein manifold, calculated using
the results of \cite{MSY, MSY2}, is greater than that of the round sphere,
then the metric cannot exist.
\end{itshape}
\end{quote}
It is not immediately obvious that
this can ever happen, but we shall see later that this remarkably simple
fact can often serve as a powerful obstruction. We will also discuss the
AdS/CFT interpretation of this result.

The Reeb vector field $\xi$
also leads to a second possible obstruction.
Given $\xi$ 
for a putative Sasaki--Einstein metric, it 
is a simple matter to show that holomorphic functions $f$ on the 
corresponding cone $X$ with definite charge $\lambda>0$, 
\be 
\mathcal{L}_{\xi}
f = \lambda i f
\ee 
give rise to
eigenfunctions of the Laplacian on the Sasaki--Einstein manifold 
with eigenvalue $\lambda(\lambda+2n-2)$.
Lichnerowicz's theorem \cite{lich} states that the \emph{smallest} 
eigenvalue of this Laplacian is bounded from below by the 
dimension of the manifold, and this
leads to the restriction $\lambda\ge 1$. Thus we have what we will call
the {\bf Lichnerowicz obstruction:}
\begin{quote}
\begin{itshape}
   If one can demonstrate the existence
of a holomorphic function on $X$ with positive charge $\lambda<1$ with respect to the 
putative Reeb vector
 field $\xi$, 
one concludes that no Sasaki--Einstein metric can exist with this Reeb 
vector field. 
\end{itshape}
\end{quote}
Again, it is not immediately obvious that this 
can ever happen. Indeed, if $\xi$ is regular, so that the orbit 
space $V$ is a Fano manifold, we show that 
this cannot happen. Nevertheless, there 
are infinitely many examples of simple hypersurface singularities 
with non--regular Reeb vector fields that violate Lichnerowicz's bound.

We shall show that for Calabi--Yau 3--folds and 4--folds, the Lichnerowicz obstruction has a beautiful
AdS/CFT interpretation: holomorphic functions on the cone $X$ are dual to
chiral primary operators in the dual superconformal field theory. 
The Lichnerowicz bound then translates into the unitarity bound for the dimensions of the operators.

The plan of the rest of the paper is as follows. In section \ref{section2} we discuss the two
obstructions in a little more detail. In section \ref{section3} we investigate the obstructions in the context of
isolated quasi--homogeneous hypersurface singularities. We also compare our results with the 
sufficient conditions reviewed in \cite{BGreview} for existence 
of Sasaki--Einstein metrics on links of such singularities. In section \ref{3fold} we show that
some 3--fold examples discussed in \cite{vafa} do not admit
Ricci--flat K\"ahler cone metrics. We briefly discuss the implications for the dual field theory.
The results of this section leave open the possibility\footnote{Recently 
reference \cite{Conti} appeared. The conclusions of the latter 
imply that this solution does not in fact exist.} of a single new 
cohomogeneity one Sasaki--Einstein 
metric on $S^5$ and we present some details of the relevant 
ODE that needs to be solved in an appendix.
In section \ref{section5} we show that some of the 4--fold examples discussed in \cite{GVW}
also do not admit Ricci--flat K\"ahler cone metrics. 
Section \ref{concl} briefly concludes.

\section{The obstructions}
\label{section2}

In this section we describe two obstructions to the 
existence of a putative Sasaki--Einstein metric on the link of an isolated Gorenstein 
singularity $X$ with Reeb vector field $\xi$. These are 
based on Bishop's theorem \cite{bishop} and Lichnerowicz's theorem 
\cite{lich}, respectively. 
We prove that the case when $\xi$ generates
a freely acting circle action, with orbit space a Fano manifold $V$, 
is never obstructed by Lichnerowicz. We also give an interpretation of Lichnerowicz's bound
in terms of the unitarity bound in field theory, via the AdS/CFT correspondence.

Let $X$ be an isolated Gorenstein singularity, and $X_0$ be the smooth 
part of $X$. We take $X_0$ to be diffeomorphic as 
a real manifold to $\R_+\times L$ where $L$ is compact, and 
let $r$ be a coordinate
on $\R_+$ with $r>0$, so that $r=0$ is the isolated singular point of $X$.  
We shall refer to $L$ as the link of the singularity. 
Since $X$ is Gorenstein, by definition there exists a nowhere zero 
holomorphic $(n,0)$--form $\Omega$ on $X_0$. 

Suppose that $X$ admits a K\"ahler metric that is a cone 
with respect to a homothetic vector field $r\partial/\partial r$, as in
(\ref{cone}). This in particular means that $L$ 
is the orbit space of $r\partial/\partial r$ and $g_L$ is a Sasakian metric. 
The Reeb vector field is defined to be
\be
\xi = J\left(r\frac{\partial}{\partial r}\right)~.\ee
In the special case that the K\"ahler metric on $X$ is Ricci--flat, 
the case of central interest, $(L,g_L)$ is Sasaki--Einstein and we have
\be\label{cons}
\mathcal{L}_{\xi}\Omega = ni\Omega\ee
since $\Omega$ is homogeneous of degree $n$ under $r\partial/\partial r$. 
This fixes the normalisation of $\xi$.

\subsection{The Bishop obstruction}

The volume $\vol(L,g_L)$ of a Sasakian metric on the
link $L$ depends only on the Reeb vector field \cite{MSY2}. 
Thus, specifying a Reeb vector field $\xi$ for a putative Sasaki--Einstein 
metric on $L$ is sufficient to specify the volume, 
assuming that the metric in fact exists. We define the 
normalised volume as
\be
V(\xi) = \frac{\vol(L,g_L)}{\vol(S^{2n-1})}
\ee
where $\vol(S^{2n-1})$ is the volume of the round sphere. 
Since Bishop's theorem \cite{bishop} (see also \cite{Besse}) implies 
that for any $(2n-1)$--dimensional Einstein manifold $(L,g_L)$ 
with $\mathrm{Ric}=2(n-1)g_L$
\be
\vol(L,g_L)\leq \vol(S^{2n-1})\ee
we immediatley have
\begin{quote}{\bf Bishop obstruction:}
\begin{itshape}
   Let $(X, \Omega)$ be an isolated Gorenstein singularity with link $L$ and
putative Reeb vector field $\xi$. If $V(\xi)>1$ then $X$ admits no 
Ricci--flat K\"ahler cone metric with Reeb vector field $\xi$.
In particular $L$ does not admit a Sasaki--Einstein metric with this 
Reeb vector field.
\end{itshape}
\end{quote}
There are a number of methods for computing the normalised volume $V(\xi)$. 
For quasi--regular $\xi$, the volume $V(\xi)$ is essentially 
just a Chern number, which makes it clear that $V(\xi)$ is a 
holomorphic invariant. In general, one can compute $V(\xi)$ as a 
function of $\xi$, and a number of different formulae have been 
derived in \cite{MSY, MSY2}. In \cite{MSY2} a general formula 
for the normalised volume $V(\xi)$ was given that involves 
(partially) resolving the singularity $X$ and applying localisation. 
For toric Sasakian manifolds
there is a simpler formula 
\cite{MSY}, giving
the volume in terms of the toric data defining the singularity.
In this paper we shall instead exploit the fact that the volume $V(\xi)$ can be
extracted from a limit of a certain index--character \cite{MSY2}; this 
is easily computed algebraically for isolated hypersurface 
singularities, which shall constitute our main set of examples in this paper.
We briefly recall some of the details from \cite{MSY2}.
Suppose we have a holomorphic $(\C^*)^r$ action on $X$. 
We may define the character
\be\label{char}
C({\bf q},X) = \mathrm{Tr} \ {\bf q}\ee
as the trace\footnote{As in \cite{MSY2}, we don't worry 
about where this trace converges, since we are mainly interested in 
the behaviour near a certain pole.} of the action of 
${\bf q}\in (\C^*)^r$ on the holomorphic 
functions on $X$. Holomorphic functions $f$ on $X$ that are 
eigenvectors of the induced $(\C^*)^r$ action 
\be
(\C^*)^r:\, f \to \mathbf{q}^{\mathbf{m}} f~,
\ee
with eigenvalue $\mathbf{q}^{\mathbf{m}}=
\prod_{a=1}^r q_a^{m_a}$ 
form a vector space over $\C$ 
of dimension $n_{\mathbf{m}}$. Each eigenvalue then contributes 
$n_{\mathbf{m}} \mathbf{q}^{\mathbf{m}}$ to the trace (\ref{char}).
Let $\zeta_a$ form a basis for 
the Lie algebra of $U(1)^r\subset (\C^*)^r$, and write the Reeb vector field
as
\be
\xi = \sum_{a=1}^r b_a \zeta_a~.\ee
Then the volume of a Sasakian metric on $L$ with Reeb vector field 
$\xi$, relative to that of the round sphere, is given by
\be\label{MSYlimit}
V(\xi)= \lim_{t\rightarrow 0} \ t^n \ C(q_a = \exp(-tb_a),X)~.\ee
In general, the right hand side of this formula may 
be computed by partially resolving $X$ and using localisation. However, for 
isolated quasi--homogeneous hypersurface singularities  
it is straightforward to compute this algebraically.

In addition, it was shown in \cite{MSY2} that the Reeb vector field 
for a Sasaki--Einstein metric on $L$ extremises $V$
as a function of the $b_a$, subject to the constraint (\ref{cons}).
This is a geometric analogue of $a$--maximisation \cite{IW} in 
superconformal field theories.

\subsection{The Lichnerowicz obstruction}
\label{lichsec}

Let $f$ be a holomorphic function on $X$ with 
\be
\mathcal{L}_{\xi} f = \lambda if\ee
where $\R\ni \lambda>0$, 
and we refer to $\lambda$ as the charge of $f$ under 
$\xi$. 
Since $f$ is holomorphic, this immediately implies that
\be
f = r^{\lambda}\tilde{f}\ee
where $\tilde{f}$ is homogeneous degree zero under $r\partial/\partial r$ 
-- that is, $\tilde{f}$ is the pull--back to $X$ 
of a function on the link $L$. 
Moreover, since $(X,g_X)$ is K\"ahler,
\be
\nabla^2_X f = 0\ee
where $-\nabla^2_X$ is the Laplacian on $(X,g_X)$. For a 
metric cone, this is related to the Laplacian 
on the link $(L,g_L)$ at $r=1$ by
\be
\nabla^2_X = \frac{1}{r^2}\nabla^2_L +\frac{1}{r^{2n-1}}\frac{\partial}{
\partial r}\left(r^{2n-1}\frac{\partial}{\partial r}\right)~.\ee
From this, one sees that
\be
-\nabla^2_L \tilde{f} = E\tilde{f}\ee
where
\be\label{energy}
E = \lambda[\lambda+(2n-2)]~.\ee
Thus any holomorphic function $f$ of definite charge under $\xi$, or 
equivalently degree under $r\partial/\partial r$, 
corresponds to an eigenfunction of the Laplacian on the link. 
The charge $\lambda$ is then related simply to the eigenvalue $E$ by the above 
formula (\ref{energy}). 

By assumption, $(X,g_X)$ is Ricci--flat K\"ahler, which implies
 that $(L,g_L)$ is Einstein with 
Ricci curvature $2n-2$. The first non--zero 
eigenvalue $E_1>0$ of $-\nabla^2_L$ is bounded from below:
\be
E_1 \geq 2n-1~.\ee
This is Lichnerowicz's theorem \cite{lich}. 
Moreover, equality holds if and only 
if $(L,g_L)$ is isometric to the round sphere $S^{2n-1}$  \cite{obata}. 
This is important as we shall find examples of links, that 
are not even diffeomorphic to the sphere, which hit this bound.
From (\ref{energy}), we immediately see that Lichnerowicz's 
bound becomes $\lambda\geq 1$.

This leads to a potential holomorphic obstruction to 
the existence of Sasaki--Einstein metrics:
\begin{quote}{\bf Lichnerowicz obstruction:}
\begin{itshape}
   Let $(X, \Omega)$ be an isolated Gorenstein singularity with link $L$ and
putative Reeb vector field $\xi$. Suppose that there 
exists a holomorphic function $f$ on $X$ of positive charge 
$\lambda<1$ under $\xi$. Then $X$ admits no Ricci--flat K\"ahler cone metric with 
Reeb vector field $\xi$. In particular $L$ does not admit a 
Sasaki--Einstein metric with this Reeb vector field.
\end{itshape}
\end{quote}
As we stated earlier, it is not immediately clear that this can 
ever happen. In fact, there are examples of 
hypersurface singularities where this 
serves as the only obvious simple 
obstruction, as we explain later. However, in the 
next subsection we treat a situation where Lichnerowicz never obstructs.

Before concluding this subsection we note that the volume 
of a Sasakian metric on $L$ with Reeb vector field $\xi$ is also 
related to holomorphic functions on $X$ of definite charge, 
as we briefly reviewed in the previous subsection. In fact 
we may write (\ref{MSYlimit}) as
\be\label{limit}
V(\xi) = \lim_{t\rightarrow 0}\ t^n \ \mathrm{Tr} \ \exp(-t \mathcal{L}_{r\partial/\partial r})\ee
where $r\partial/\partial r=-J(\xi)$. Here the 
trace denotes a trace of the action of 
$\mathcal{L}_{r\partial/\partial r}$ on the holomorphic functions on $X$. 
Thus a holomorphic function $f$ of charge $\lambda$ under $\xi$ contributes 
$\exp(-t\lambda)$ to the trace.  
That (\ref{limit}) agrees with (\ref{MSYlimit}) follows from the fact that
we can write $\lambda = (\mathbf{b},\mathbf{m})$.
Given our earlier discussion relating $\lambda$ to 
eigenvalues of the Laplacian on $L$, the above trace very much
resembles the trace of the heat kernel, also known as the 
partition function, on $L$. In fact, since it is a sum over 
only holomorphic eigenvalues, we propose to call it the holomorphic 
partition function. The fact that the volume of a Riemannian manifold 
appears as a pole
in the heat kernel is well known \cite{spelling}, and (\ref{limit}) 
can be considered a holomorphic Sasakian analogue.

Notice then that the Lichnerowicz obstruction involves 
holomorphic functions on $X$ of small charge with respect to $\xi$,
whereas the Bishop obstruction is a statement about the volume, which 
is determined by the asymptotic growth of holomorphic functions on $X$.

\subsection{Smooth Fanos}

Let $V$ be a smooth Fano K\"ahler manifold. Let $K$ denote the canonical line 
bundle over $V$. By definition, $K^{-1}$ is an ample holomorphic line bundle, 
which thus specifies a positive class 
\be
c_1(K^{-1})=-c_1(K)\in H^2(V;\Z)\cap H^{1,1}(V;\R)\cong\mathrm{Pic}(V) ~.\ee
Recall here that $\mathrm{Pic}(V)$ is the group of holomorphic 
line bundles on $V$. 
Let $I(V)$ denote the largest positive integer such that 
$c_1(K^{-1})/I(V)$ is an integral class in $\mathrm{Pic}(V)$.
$I(V)$ is called the \emph{Fano index} of $V$. For example, 
$I(\mathbb{CP}^2)=3$, $I(\mathbb{CP}^1\times\mathbb{CP}^1)=2$, 
$I(\mathbb{F}_1)=1$. Let $\mathcal{L}$ be the holomorphic line bundle
$\mathcal{L}=K^{1/I(V)}$, which is primitive in $\mathrm{Pic}(V)$ 
by construction. Denote the total space of 
the unit circle bundle in $\mathcal{L}$ by $L$ -- this is our link. 
We thus have a circle bundle
\be\label{fibra}
S^1\hookrightarrow L\rightarrow V\ee
where $\mathcal{L}$ is the associated line bundle. 
If $V$ is simply--connected then $L$ is also simply--connected, 
as follows from the Gysin sequence of the 
fibration (\ref{fibra}). 
Note that $V$ admits a K\"ahler--Einstein metric if and only if $L$ 
admits a regular Sasaki--Einstein metric with Reeb vector field 
that rotates the $S^1$ fibre of (\ref{fibra}). 

$X$ is obtained from the total space of $\mathcal{L}$ by collapsing 
(or deleting, to obtain $X_0$) the zero section. Holomorphic functions on $X$
of definite charge 
are then in 1--1 correspondence with global 
sections of $\mathcal{L}^{-k}$, which are elements of the group 
$H^0(\mathcal{O}(\mathcal{L}^{-k}))$. 
Let $\zeta$ be the holomorphic vector field on $X$ that rotates the fibre 
of $\mathcal{L}$ with weight one. That is, if $s\in H^0(\mathcal{O}(\mathcal{L}^{-1}))$ is 
a holomorphic section of the ample line bundle $\mathcal{L}^{-1}$, 
viewed as a holomorphic function on $X$, then
\be
\mathcal{L}_{\zeta}s=is~.\ee
Since $K=\mathcal{L}^{I(V)}$ is the canonical bundle of $V$, it follows that
the correctly normalised Reeb vector field is (see, for example, \cite{MSY2})
\be
\xi = \frac{n}{I(V)}\zeta~.\ee
We briefly recall why this is true. Let $\psi$ be a local coordinate 
such that $\xi=\partial/\partial\psi$. Then $n\psi/I(V)$ is a 
local coordinate on the circle fibre of (\ref{fibra}) with period 
$2\pi$. This follows since locally the contact one--form of 
the Sasakian manifold is
$\eta=\diff\psi - A$, where $A/n$ is a connection on the 
canonical bundle of $V$. 

Holomorphic functions of smallest positive charge obviously 
correspond to $k=1$. 
Any section $s\in H^0(\mathcal{O}(\mathcal{L}^{-1}))$ then has charge
\be
\lambda = \frac{n}{I(V)}\ee
under $\xi$. However, it is well known 
(see, for example, \cite{Kollarbook}, page 245) that for smooth 
Fanos $V$ we have $I(V)\leq n$, with $I(V)=n$ 
if and only if $V=\mathbb{CP}^{n-1}$. Thus, in this 
situation, we always have $\lambda\geq1$ and Lichnerowicz
never obstructs. 
\begin{quote}
\begin{itshape}
Lichnerowicz's theorem  can only obstruct 
for non--regular Reeb vector fields. 
\end{itshape}\end{quote}
We expect a similar statement to be true for 
the Bishop bound. For a regular Sasaki--Einstein manifold with
Reeb vector field $\xi$ and
orbit space a Fano manifold $V$, Bishop's bound may be written
\be\label{top}
I(V)\int_V c_1(V)^{n-1}\leq n\int_{\mathbb{CP}^{n-1}}c_1(\mathbb{CP}^{n-1})^{n-1} = n^n~.\ee
It seems reasonable to expect the topological 
statement (\ref{top}) to be true for 
any Fano manifold $V$, so that Bishop never obstructs in the regular case, 
although we are unaware of any proof. Interestingly, this is closely 
related to a standard conjecture in algebraic geometry, that bounds 
$\int_V c_1(V)^{n-1}$ from above by $n^{n-1}$ for any Fano manifold $V$, with 
equality if and only if $V=\mathbb{CP}^{n-1}$. In general, this 
stronger statement is false (see \cite{Kollarbook}, page 251), although 
it is believed to be true in the special case that $V$ has Picard
number one, {\it i.e.} $\mathrm{rank}(\mathrm{Pic}(V))=1$. This has recently 
been proven up to dimension $n=5$ \cite{hwang}. It would be interesting 
to investigate (\ref{top}) further.

\subsection{AdS/CFT interpretation}

In this section we show that the Lichnerowicz obstruction has 
a very natural  interpretation in the AdS/CFT dual field theory, 
in terms of a unitarity bound. We also briefly discuss the Bishop bound.   

Recall that every superconformal field theory possesses 
a supergroup of symmetries and that the AdS/CFT duality maps 
this to the superisometries of the dual geometry. 
In particular, in the context of Sasaki--Einstein geometry, it maps
the R--symmetry in the field theory to the isometry generated by 
the Reeb vector field $\xi$,  and 
the R--charges of operators in the field theory are proportional to the weights under $\xi$.
Generically, Kaluza--Klein excitations in the geometry correspond to
gauge invariant operators in the field theory.
 These operators are characterised by their scaling dimensions $\Delta$. 
 The supersymmetry algebra then implies that a
  general operator satisfies a BPS bound relating the dimension  
 to the R--charge $R$: 
 $\Delta\geq (d-1)R/2 $. When this bound is saturated the corresponding BPS operators belong to 
 short representations of the supersymmetry algebra, and in particular are chiral.  
  Here we will only consider scalar gauge invariant operators which are chiral.

It is well known that for any conformal field theory, 
in arbitrary dimension $d$, 
the scaling dimensions of all operators are bounded as a consequence of unitarity. 
In particular, for scalar operators, we have
\be
\Delta\geq \frac{d-2}{2}~.
\label{unibound}
\ee
In section \ref{lichsec} we have argued that a necessary condition 
for the existence of a
Sasaki--Einstein metric is that 
the charge $\lambda>0$ of any holomorphic function 
on the corresponding Calabi--Yau cone must satisfy the bound 
\be
\lambda\geq 1~.
\ee 
In the following, we will show that these two bounds coincide.

We start with a gauge theory realised on the world--volume of a large
number of D3 branes, placed at a 3--fold Gorenstein singularity $X$.
The affine variety $X$ can then be thought of as (part of)
the moduli space of vacua of this gauge theory.
In particular, the holomorphic functions, defining the
coordinate ring of $X$, correspond to 
(scalar) elements of the chiral ring of the gauge theory \cite{nekrasov2}.
Recalling that an AdS$_{4/5}\times L^{7/5}$ solution arises as the 
near--horizon limit of
a large number of branes at a Calabi--Yau 4--fold/3--fold conical singularity,
it is clear that the weights $\lambda$ of these holomorphic 
functions under the action 
of $r\de /\de r$ must be proportional to the scaling dimensions $\Delta$ of 
the dual operators, corresponding to excitations in AdS space. 
We now make this relation more precise.

According to the AdS/CFT dictionary \cite{Gubser:1998bc,Witten:1998qj}, 
a generic scalar excitation $\Phi$ in AdS obeying
\be
(\Box_{\mathrm{AdS}_{d+1}}-m^2)\,\Phi=0
\ee
and which behaves like $\rho^{-\Delta}$ near the boundary of AdS
($\rho\to \infty$), is dual to an operator in the dual CFT with scaling dimension
\be\label{old}
m^2=\Delta (\Delta-d)\quad \Rightarrow\quad
\Delta_\pm = \frac{d}{2}\pm \sqrt{\frac{d^2}{4}+m^2}~.
\ee 
More precisely, for $m^2\ge -d^2/4+1$ the dimension of the operator 
is given by $\Delta_+$. However, for $-d^2/4<m^2<-d^2/4+1$ one can take
either $\Delta_\pm$ and these will correspond to inequivalent CFTs \cite{KW2,Balasubramanian:1998sn}. 
Notice that $\Delta_+$ is always well above the bound implied by unitarity.
On the other hand, $\Delta_-$ saturates this bound for $m^2=-d^2/4+1$.  

The values for $m^2$ can be obtained from the eigenvalues $E$ of the scalar Laplacian 
$-\nabla^2_L$ on the internal manifold $L$ by performing a Kaluza--Klein analysis. 
The modes corresponding to the chiral primary operators have been identified
in the literature in the context of a more general analysis for Einstein manifolds; see 
\cite{KW2,ceresole} for type IIB supergravity compactified on $L^5$, and \cite{fabbri,fabbri2} 
for M--theory compactified on $L^7$.

Consider first $d=4$ ({\it i.e.} $n=3$).
The supergravity modes dual to chiral operators are 
a mixture of the trace mode of the internal metric and the RR four--form and lead to 
\cite{Kim,KW2,ceresole} 
\be
m^2 = E+16-8\sqrt{E+4}~.
\ee  
Combining this with (\ref{energy}) it follows that
\be
\Delta_\pm = 2\pm |\lambda-2|
\ee
so that $\Delta=\lambda$,
providing that we take $\Delta_-$ for $\lambda<2$ and $\Delta_+$ for $\lambda\ge 2$. Notice that for 
$\lambda=2$, $\Delta_+=\Delta_-$, and this corresponds to the 
Breitenlohner--Freedman bound $m^2(\lambda=2)=-4$ 
for stability in AdS$_5$.

The case $d=3$ ({\it i.e.} $n=4$), relevant for AdS$_4\times L^7$ geometries, is similar.
The scalar supergravity modes corresponding to chiral primaries \cite{fabbri,fabbri2} are 
again a mixture of the metric trace and the three--form potential \cite{castellani},
and fall into short ${\cal N}=2$ multiplets. Their masses are given 
by\footnote{Note that the mass formulae in \cite{dauria} are relative to the 
 operator $\Box_{AdS_4}-32$. Moreover, the factor of four mismatch
  between their  $m^2$ and ours is simply due to the fact that it is actually 
  $m^2R^2$ that enters in (\ref{old}), and the radius of AdS$_4$ is $1/2$ that of AdS$_5$.} \cite{castellani,dauria,fabbri2} 
\be
m^2 = \frac{E}{4}+9 -3\sqrt{E+9}~.
\ee
Combing this with (\ref{energy}) it follows that
\be
\Delta_\pm=\frac{1}{2}\left(3\pm|\lambda-3|\right)
\ee
so that $\Delta=\frac{1}{2}\lambda$,
providing that we take $\Delta_-$ for $\lambda<3$ and $\Delta_+$ for $\lambda\ge 3$. Once again
the switching of the two branches occurs at the Breitenlohner--Freedman 
bound $m^2(\lambda=3)=-9/4$ 
for stability in AdS$_4$. 

In summary, we have shown that 
\bea
\Delta=\left\{
\begin{array}{cc}
\lambda & \qquad\mathrm{for}~d=4\\
\frac{1}{2}\lambda &\qquad\mathrm{for}~d=3
\end{array}
\right.~.
\eea
Thus in both cases relevant for AdS/CFT the Lichnerowicz  bound $\lambda\geq 1$ is 
equivalent to the unitarity bound (\ref{unibound}).

The Bishop bound also has a direct interpretation 
in field theory. 
Recall that 
the volume of the Einstein 5--manifold $(L,g_L)$ is related to the 
exact $a$ central charge 
of the dual four dimensional conformal field theory via \cite{kostas}
(see also \cite{Gubser})
\be\label{barry}
a(L) = \frac{\pi^3 N^2}{4\vol(L,g_L)}
\ee
where $N$ is the number of D3--branes. The Bishop bound then implies that 
\be\label{inq}
a(L) \ge \frac{N^2}{4} = a({\cal N}=4) 
\ee
where $N^2/4$ is the central charge of ${\cal N}=4$ super Yang--Mills theory. 
One can give a heuristic argument for this inequality, as 
follows\footnote{We thank 
Ken Intriligator for this argument.}. 
By appropriately Higgsing the dual field theory, and then integrating out the massive
fields, one expects to be able to flow to ${\cal N}=4$ super 
Yang--Mills theory. This is because the Higgsing corresponds to moving the
D3--branes away from the singular point to a smooth point of the cone, at which
the near horizon geometry becomes $AdS_5\times S^5$. 
Since the number of massless degrees of freedom is expected to decrease in such a process, 
we also expect that the $a$ central charge to decrease. 
This would then explain the inequality (\ref{inq}).

\section{Isolated hypersurface singularities}
\label{section3}

In this section we describe links of 
isolated quasi--homogeneous 
hypersurface singularities. These provide many simple 
examples of both obstructions. 

Let $w_i\in \Z_+$, $i=1,\ldots, n+1$, be a set of positive weights. 
We denote these by a vector $\mathbf{w}\in (\Z_+)^{n+1}$.
This defines an action of $\C^*$ on $\C^{n+1}$ via
\be
(z_1,\ldots,z_{n+1})\mapsto (q^{w_1}z_1,\ldots,q^{w_{n+1}}z_{n+1})\ee
where $q\in\C^*$. Without loss of generality one can take 
the set $\{w_i\}$ to have no common factor. This ensures 
that the above $\C^*$ action is effective. However, for 
the most part, this is unnecessary for our purposes 
and we shall not always do this. Let
\be
F:\C^{n+1}\rightarrow \C\ee
be a quasi--homogeneous polynomial on $\C^{n+1}$ with 
respect to $\mathbf{w}$. This means 
that $F$ has definite degree $d$ under the above $\C^*$ action:
\be
F(q^{w_1}z_1,\ldots,q^{w_{n+1}}z_{n+1})=q^d 
F(z_1,\ldots,z_{n+1})~.\ee 
Moreover we assume that the affine algebraic variety
\be
X=\{F=0\}\subset \C^{n+1}\ee
is smooth everywhere except at the origin $(0,0,\ldots,0)$. 
For obvious reasons, such $X$ are called isolated 
quasi--homogeneous hypersurface singularities. 
The corresponding link $L$ is the
intersection of $X$ with the unit sphere in $\C^{n+1}$:
\be
\sum_{i=1}^{n+1}|z_i|^2=1~.
\ee

A particularly nice set of such singularities are provided by 
so--called Brieskorn--Pham singularities. These take the particular 
form
\be\label{BP}
F = \sum_{i=1}^{n+1} z_i^{a_i}\ee
with $\mathbf{a}\in (\Z_+)^{n+1}$. Thus the weights of the $\C^*$ action are given by 
 $w_i=d/a_i$. 
The corresponding hypersurface singularities $X$ are always isolated, as
is easily checked. Moreover, the topology of the links $L$ are also 
extremely well understood -- see \cite{Randell} for a complete 
description of the homology groups of $L$. In particular, 
$L$ is known to be $(n-2)$--connected, meaning that the homotopy 
groups are $\pi_a(L)=0$ for 
all $a=1,\ldots,n-2$. 

Returning to the general case, we may define a nowhere zero holomorphic $(n,0)$--form $\Omega$ 
on the smooth part of $X$ by
\be
\Omega = \frac{\diff z_1\wedge\cdots\wedge\diff z_n}{\partial F/\partial z_{n+1}}~.\ee
This defines $\Omega$ on the patch where $\partial F/\partial z_{n+1}\neq0$. One has 
similar expressions on patches where $\partial F/\partial z_i\neq0$ for each $i$, and it is simple to 
check that these glue together into a nowhere zero form $\Omega$. 
Thus all such $X$ are Gorenstein, and moreover they come equipped 
with a holomorphic $\C^*$ action by construction. The orbit space of this 
$\C^*$ action, or 
equivalently the orbit space of $U(1)\subset \C^*$ on the link, is 
a complex orbifold $V$. In fact, $V$ is the weighted 
variety defined by $\{F=0\}$ in the weighted
projective space $\mathbb{WCP}^n_{[w_1,w_2,\ldots,w_{n+1}]}$. 
The latter is the quotient of the non--zero vectors in $\C^{n+1}$ by 
the weighted $\C^*$ action
\be
\mathbb{WCP}^n_{[w_1,w_2,\ldots,w_{n+1}]}=\left(\C^{n+1}\setminus \{(0,0,\ldots,0)\}\right)/\C^*\ee
and is a complex orbifold with a natural K\"ahler orbifold 
metric, up to scale, induced from K\"ahler reduction 
of the flat metric on $\C^{n+1}$. 
It is not difficult to show that $V$ is a Fano orbifold if and only if 
\be
|\mathbf{w}|-d>0\ee
where $|\mathbf{w}|=\sum_{i=1}^{n+1}w_i$. 
To see this, first notice that 
$|\mathbf{w}|-d$ is the charge of $\Omega$ under $U(1)\subset \C^*$. 
To be precise, if $\zeta$ denotes the holomorphic vector field on $X$ with
\be
\mathcal{L}_{\zeta}z_j = w_j i z_j\ee
for each $j=1,\ldots,n+1$, then
\be
\mathcal{L}_{\zeta}\Omega = (|\mathbf{w}|-d) i\Omega~.\ee
Positivity of this charge $|\mathbf{w}|-d$ then implies \cite{MSY2} that 
the cohomology class of the natural Ricci--form induced on $V$ is represented by 
a positive $(1,1)$--form, which is the definition that $V$ is Fano.

If there exists a Ricci--flat K\"ahler metric on $X$ which is a cone 
under $\R_+\subset \C^*$, then the correctly normalised 
Reeb vector field is thus
\be
\xi = \frac{n}{|\mathbf{w}|-d}\zeta~.\ee
We emphasise that here we will focus on the possible (non--)existence of a
Sasaki--Einstein metric on the link $L$ which has this 
canonical vector field as its Reeb vector field. It is 
possible that such
metrics are obstructed, but that there exists a Sasaki--Einstein metric
on $L$ with a different Reeb vector field. This may be investigated 
using the results of \cite{MSY2}. In particular we 
shall come back to this point for a class of 3--fold examples in 
section \ref{3fold}.

\subsection{The Bishop obstruction} 

A general formula for the volume of a Sasaki--Einstein metric 
on the link of an  isolated quasi--homogeneous 
hypersurface singularity was given in 
\cite{BH}. Strictly speaking, this formula was proven only 
when the Fano $V$ is \emph{well--formed}. This means that the orbifold 
loci of $V$ are at least complex codimension two. When $V$ is not 
well--formed, the singular sets of $V$ considered as an orbifold and as an algebraic 
variety are in fact different. A simple example is 
the weighted projective space $\mathbb{WCP}^1_{[p,q]}$ where 
$\mathrm{hcf}(p,q)=1$. As an orbifold, this is topologically 
a 2--sphere with conical singularities at the north and south poles 
of polar angle $2\pi/p$ and $2\pi/q$, respectively. As an algebraic 
variety, this weighted projective space is just $\mathbb{CP}^1$ since $\mathbb{C}/\Z_p=\C$. 
In fact, as a manifold it is diffeomorphic to $S^2$, for the same 
reason. When we say K\"ahler--Einstein orbifold metric, we must 
keep track of this complex codimension one orbifold data in the 
non--well--formed case. For further 
details, the reader is directed to the review \cite{BGreview}.

Assuming that there exists a Sasaki--Einstein 
metric with Reeb action $U(1)\subset\C^*$, then the volume of this 
link  when $V$ is well--formed is given by \cite{BH}
\be\label{BHformula}
\mathrm{vol}(L) = \frac{2d}{w(n-1)!}\left(\frac{\pi(|\mathbf{w}|-d)}{n}\right)^n~.\ee
Here $w=\prod_{i=1}^{n+1}w_i$ denotes the product of the weights. 

Using the earlier formula (\ref{MSYlimit}), we may now give an 
alternative derivation of this formula. The advantage of this approach 
is that, in contrast to \cite{BH}, we never descend to the orbifold $V$. 
This allows us to dispense with the well--formed condition, and 
show that (\ref{BHformula}) holds in general. The authors of \cite{BH} 
noted that their formula seemed to apply to the general case.

Let us apply (\ref{MSYlimit}) to isolated quasi--homogeneous hypersurface 
singularities. Let $q\in \C^*$ denote the weighted action on $X$. 
We may compute the 
character $C(q,X)$ rather easily, since holomorphic functions 
on $X$ descend from holomorphic functions on $\C^{n+1}$, 
and the trace over the latter is simple to compute. A discussion of
precisely this problem may be found in \cite{Nekrasov}. According 
to the latter reference, the character is simply
\be
C(q,X) = \frac{1-q^d}{\prod_{i=1}^{n+1}(1-q^{w_i})}~.\ee
The limit (\ref{MSYlimit}) is straightforward to take, giving the 
normalised volume
\be
V(\xi) = \frac{d}{wb^n}\ee
where, as above,
\be
\xi = b\zeta\ee
and $\zeta$ generates the $U(1)\subset\C^*$ action. Thus, from our 
earlier discussion on the charge of $\Omega$, we have
\be
b = \frac{n}{|\mathbf{w}|-d}\ee
giving
\be
\vol(L) = \frac{d\left(|\mathbf{w}|-d\right)^n}{wn^n}\vol(S^{2n-1})~.\ee
Restoring
\be
\vol(S^{2n-1}) = \frac{2\pi^n}{(n-1)!}\ee
we thus obtain the result (\ref{BHformula}).

Bishop's theorem then requires, for existence of a Sasaki--Einstein 
metric on $L$ with Reeb vector field $\xi$ generating the canonical $U(1)$ action, 
\be\label{bishbound}
d\left(|\mathbf{w}|-d\right)^n\leq wn^n~.\ee
We shall see that infinitely many isolated quasi--homogeneous
hypersurface singularities with Fano $V$ violate this 
inequality.

\subsection{The Lichnerowicz obstruction}

As we already mentioned, holomorphic functions on $X$ are simply restrictions of holomorphic 
functions on $\C^{n+1}$. Thus the smallest positive charge 
holomorphic function is $z_m$, where $m\in\{1,\ldots,n+1\}$ is such that
\be
w_m = \min \{w_i, i=1,\ldots,n+1\}~.\ee
Of course, $m$ might not be unique, but this is irrelevant since all such
$z_m$ have the same charge in any case. This charge is
\be
\lambda = \frac{nw_m}{|\mathbf{w}|-d}\ee
and thus the Lichnerowicz obstruction becomes
\be\label{lichbound}
|\mathbf{w}|-d\leq nw_m~.
\ee
Moreover, this bound can be saturated if and only if $X$ is $\C^{n}$ with 
its flat metric. 

It is again clearly trivial to construct many 
examples of isolated hypersurface 
singularities that violate this bound. 

\subsection{Sufficient conditions for existence}

In a series of works by Boyer, Galicki and collaborators, many 
examples of Sasaki--Einstein metrics have been shown to exist on 
links of isolated quasi--homogeneous hypersurface singularities of the form
(\ref{BP}).
Weighted homogeneous perturbations of these singularities can lead 
to continuous families of Sasaki--Einstein metrics. For a recent review 
of this work, we refer the reader to \cite{BGreview} and references 
therein. 

Existence of these metrics is proven using the continuity method. 
One of the sufficient (but far from necessary) conditions for there 
to exist a Sasaki--Einstein metric is that the weights 
satisfy the condition \cite{BGreview} 
\be\label{zippy}
|\mathbf{w}| - d <\frac{n}{(n-1)}w_m~.\ee
In particular, for $n>2$ this implies that
\be
|\mathbf{w}| - d < nw_m\ee
which is precisely Lichnerowicz's bound. 
Curiously, for $n=2$ the Lichnerowicz bound and (\ref{zippy})
are the same, although this case is rather trivial.


\section{A class of 3--fold examples}
\label{3fold}

Our first set of examples are given by the 3--fold singularities 
with weights $\mathbf{w}=(k,k,k,2)$ and polynomial
\be
F = \sum_{i=1}^3 z_i^2 + z_4^k\ee
where $k$ is a positive integer. The corresponding isolated 
hypersurface singularities $X_k=\{F=0\}$ are of Brieskorn--Pham type.
Notice that 
$X_1=\C^3$ and $X_2$ is the ordinary double point singularity, 
better known to physicists as the conifold. Clearly, both of these admit
Ricci--flat K\"ahler cone metrics and moreover, the Sasaki--Einstein 
metrics are homogeneous. 

The differential topology of the links $L_k$ can be deduced using the 
results of \cite{Randell}, together with Smale's 
theorem for 5--manifolds.
In particular, for $k$ odd, the link $L_k$ is diffeomorphic 
to $S^5$. For $k=2p$ even, one can show
that $L_{2p}\cong S^2\times S^3$ (alternatively, see Lemma 7.1 of \cite{richard}). 

The Fanos $V_k$ are not well--formed for $k>2$. In fact, the 
subvariety $z_4=0$ in $V_k$ is a copy of $\mathbb{CP}^1$, which is 
a locus of $\Z_k$ orbifold singularities for $k$ odd, and 
$\Z_{k/2}$ orbifold singularities for $k$ even. As algebraic 
varieties, all the odd $k$ are equivalent to $\mathbb{CP}^2$, 
and all the even $k$ are equivalent to $\mathbb{CP}^1\times\mathbb{CP}^1$.
As orbifolds, they are clearly all distinct. 

\subsection{Obstructions}

These singularities
have appeared in the physics literature \cite{vafa} 
where it was assumed that all $X_k$ admitted conical Ricci--flat K\"ahler 
metrics, with Reeb action corresponding to the canonical $U(1)$ action. 
In fact, it is trivial to show that the Bishop bound 
(\ref{bishbound}) is violated 
for all $k>20$. Moreover, the Lichnerowicz bound (\ref{lichbound}) is 
even sharper: for $k\geq 2$, $z_4$ has smallest\footnote{For $k=1$, $z_i, i=1,2,3$ have 
the smallest charge $\lambda=1$. This is consistent with the fact that $k=1$ corresponds to the link 
$L_1=S^5$ with its round metric.} 
charge under $\xi$, namely 
\be
\lambda = \frac{6}{k+2}\ee
which immediately rules out all $k>4$. For $k=4$ we have $\lambda=1$. 
Recall that, according to \cite{obata}, this 
can happen if and only if $L_4$ is the round sphere. But we already 
argued that $L_4=S^2\times S^3$, which rules out $k=4$ also. 
Thus the only link that might possibly admit a Sasaki--Einstein 
metric with this $U(1)$ Reeb action, apart from $k=1,2$, is $k=3$. 
We shall return to the $k=3$ case in the next subsection.

Given the contradiction, one might think that perhaps the 
canonical $\C^*$ action is not the critical one, in the sense of 
\cite{MSY2}. Writing 
\be
F = z_1^2 + uv + z_4^k\ee
there is clearly a $(\C^*)^2$ action generated by weights $(k,k,k,2)$ 
and $(0,1,-1,0)$ on $(z_1,u,v,z_4)$, respectively. The second 
$U(1)\subset \C^*$ is the maximal torus of $SO(3)$ acting on the $z_i$, $i=1,2,3$, in the 
vector representation. It is then straightforward to compute 
the volume of the link as a function of the Reeb vector field
\be
\xi = \sum_{a=1}^2 b_a\zeta_a\ee
using the character formula in \cite{Nekrasov} and taking the limit 
as in (\ref{MSYlimit}). We obtain
\be
V(b_1,b_2) = \frac{2kb_1}{2b_1\cdot kb_1 (kb_1+b_2)(kb_1-b_2)} 
= \frac{1}{b_1(kb_1+b_2)(kb_1-b_2)}~.\ee
The first component $b_1$ is fixed by the charge of $\Omega$, as above, 
to be
\be
b_1 = \frac{3}{k+2}~.\ee
According to \cite{MSY2}, the critical Reeb vector field 
for the putative Sasaki--Einstein metric is obtained by 
setting to zero the derivative of $V(b_1,b_2)$, 
with respect to $b_2$. This immediately gives $b_2=0$. 
Thus the original weighted $\C^*$ action is indeed a critical point of the 
Sasakian--Einstein--Hilbert action on the link, in this 2--dimensional 
space of Reeb vector fields.
We could have anticipated this result without computing anything. 
According to \cite{MSY2}, the critical Reeb vector field could not 
have mixed with any vector field in the Lie algebra of a 
$U(1)$ subgroup of $SO(3)$, since 
the latter group is semi--simple. 

\subsection{Cohomogeneity one metrics}

It is interesting to observe that any conical
Ricci--flat K\"ahler 
metric on $X_k$ would necessarily have 
$U(1)\times SO(3)$ isometry (the global form of the 
effectively acting isometry group will depend on $k$ mod 2).
This statement follows from 
Matsushima's theorem \cite{Matsushima}. 
Specifically, Matsushima's theorem says that the 
isometry group of a K\"ahler--Einstein manifold\footnote{The generalisation
to orbifolds is straightforward.} $(V,g_V)$ is a maximal compact subgroup of 
the group of complex automorphisms of $V$.
Quotienting $L_k$ by
the $U(1)$ action, one would thus have K\"ahler--Einstein orbifold metrics 
on $V_k$ with an $SO(3)$ isometry, whose generic orbit is three--dimensional.
In other words, these metrics, when they exist, 
can be constructed using standard 
cohomogeneity one
techniques. In fact this type of construction is very well motivated
since demanding a local $SU(2)\times U(1)^2$ isometry is one way in which 
the Sasaki--Einstein metrics of \cite{paper2} can be constructed (in 
fact they were actually found much more indirectly via M--theory \cite{paper1}). 

However, apart from the $k=3$ case, and of course the $k=1$ and $k=2$ cases, 
we have already shown that any such construction must fail. 
For $k=3$, the relevant ODEs 
that need to be solved have actually been written down in
\cite{dancer}. In appendix A we record these equations, 
as well as the boundary conditions that
need to be imposed. We have been unable 
to integrate these equations, so the question of existence
of a Sasaki--Einstein metric on $L_3$ remains open.

\subsection{Field theory}

In \cite{vafa} a family of supersymmetric quiver gauge theories 
were studied whose classical vacuum moduli space reproduces the 
affine varieties $X_{2p}$. These theories were argued to flow for large 
$N$ in 
the IR to a superconformal fixed point, AdS/CFT dual to a Sasaki--Einstein 
metric on the link $L_{2p}$ for all $p$. Indeed, the R--charges 
of fields may be computed using $a$--maximisation \cite{IW}, and 
agree with the naive geometric computations, assuming that the 
Sasaki--Einstein metrics on $L_{2p}$ exist. However, as we have 
already seen, these metrics cannot exist for any $p>1$. We argued 
in section \ref{section2} that this bound, coming from Lichnerowicz's theorem, 
is equivalent to the unitarity bound in the CFT. We indeed show 
that a gauge invariant chiral primary operator, dual to the 
holomorphic function $z_4$ that provides the geometric obstruction,
violates the unitarity bound for $p>1$. 

Before we recall the field theories for $k=2p$ even, let us 
make a remark on the $X_k$ singularities when $k$ is odd. 
In the latter case, it is not difficult to prove that 
$X_k$ admits no crepant resolution\footnote{Since the link 
$L_k=S^5$ for all odd $k$, it follows that $\mathrm{Pic}(X_k\setminus 
\{r=0\})$ is trivial, and hence $X_k$ is factorial. The isolated 
singularity at $r=0$ 
is terminal for all $k$. These two facts, together with Corollary 
4.11 of \cite{Kollar}, imply that $X_k$ has no crepant resolution.}. 
That is, there is no 
blow--up of $X_k$ to a smooth manifold $\tilde{X}$ with trivial 
canonical bundle. In such cases the field theories might be 
quite exotic, and in particular not take the form of quiver gauge theories. 
In contrast, the $X_{2p}$ singularities are resolved by 
blowing up a single exceptional $\mathbb{CP}^1$ \cite{laufer}, which 
leads to a very simple class of gauge theories.

\begin{figure}[!th]
\begin{center}
\epsfig{file=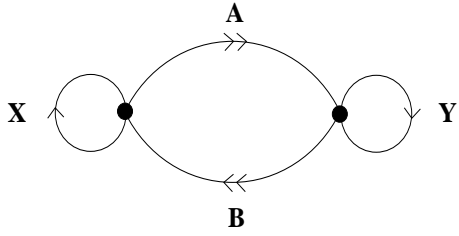,width=6cm,height=3cm}
\end{center}
\caption{Quiver diagram of the $A_1$ orbifold gauge theory.}
\label{quiver}
\end{figure}

Consider the quiver diagram for the $\mathcal{N}=2$ $A_1$ orbifold, 
depicted in Figure \ref{quiver}. 
The two nodes represent two $U(N)$ gauge groups. There are 6 matter fields:
an adjoint for each gauge group, that we denote by $X$ and $Y$, 
and two sets of bifundamental fields $A_I$ and 
$B_I$, where $I=1,2$ are $SU(2)$ flavour indices. Here the 
$A_I$ are in the $(N,\bar{N})$ representation of  $U(N)\times U(N)$, and 
the $B_I$ 
are in the $(\bar{N},N)$ representation. This is the quiver for $N$
D3--branes at the $\C\times (\C^2/\Z_2)$ singularity, 
where $\C^2/\Z_2$ is the $A_1$ surface singularity. 
However, for our field theories indexed by $p$, the superpotential is given by
\be\label{superman}
W = \mathrm{Tr}\left[X^{p+1}+(-1)^pY^{p+1}+X(A_1B_1+A_2B_2)+Y(B_1A_1+B_2A_2)\right]~.
\ee
It is straightforward to verify that the classical vacuum moduli 
space of this gauge theory gives rise to the $X_{2p}$ singularities. 
In fact these gauge theories were also studied in detail in \cite{Halmagyi}, 
and we refer the reader to this reference for further details. 
The 
$SU(2)$ flavour symmetry corresponds to the $SO(3)$ automorphism of 
$X_{2p}$. 

It is a simple matter to perform $a$--maximisation for this 
theory, taken at face value. Recall this requires one to assign 
trial R--charges to each field, and impose the constraints 
that $W$ has R--charge 2, and that the $\beta$--functions of each 
gauge group vanish. One then locally maximises the $a$--function
\be
a = \frac{3N^2}{32}\left(2+\sum_i 3(R(X_i)-1)^3 - (R(X_i)-1)\right)\ee
subject to these constraints, 
where the sum is taken over all R--charges of fields $X_i$. 
One finds the results, as in \cite{vafa}
\bea
R(X)  = R(Y) & =& \frac{2}{p+1}\nonumber\\
R(A_I) = R(B_I) & = & \frac{p}{p+1}\eea
and central charge
\be
a(L_{2p}) = \frac{27p^2N^2}{8(p+1)^3}~.\ee
This corresponds, under the AdS/CFT relation (\ref{barry}), to 
a Sasaki--Einstein volume
\be
\vol(L_{2p}) = \frac{2\pi^3(p+1)^3}{27p^2}~,
\ee
which agrees with the general formula (\ref{BHformula}). Thus an
initial reaction \cite{vafa} is that one has found agreement between geometric 
and field theory results. However, the results of this paper imply 
that the Sasaki--Einstein metrics on $L_{2p}$ \emph{do not exist} for $p>1$. 
In fact, it is clear that, upon closer inspection, the
gauge invariant  chiral primary operator $\mathrm{Tr}X$ (or $\mathrm{Tr}Y$)
has R--charge $2/(p+1)$, which violates the unitarity bound for $p>1$. In fact, 
when one computes the vacuum moduli space for a single 
D3--brane $N=1$, $\mathrm{Tr}X$ is identified 
with the holomorphic function $z_4$, and the unitarity bound and 
Lichnerowicz bound are identical, as we argued to be generally true in section 2. 

The superpotential (\ref{superman}) can be regarded as a deformation 
of the $A_1$ orbifold theory. For $p>2$, using $a$--maximisation (and assuming an $a$--theorem),
 this was argued in \cite{Halmagyi} to be an irrelevant 
deformation (rather than a ``dangerously irrelevant'' operator). 
This is therefore consistent with our geometric results. 
The case $p=2$ is interesting since it appears to be marginal. If it 
is exactly marginal, we expect a one parameter family of
solutions with fluxes that interpolates between the $A_1$ orbifold
with link $S^5/\Z_2$ and the $X_{4}$ singularity
with flux.

\section{Other examples}
\label{section5}

In this section we present some further obstructed examples. In particular 
we examine ADE 4--fold singularities, studied in \cite{GVW}. All of 
these, with the exception of $D_4$ and the obvious cases of 
$A_0$ and $A_1$, do not admit  Ricci--flat K\"ahler cone metrics 
with the canonical weighted $\C^*$ action. We 
also examine weighted $\C^*$ actions on $\C^n$.

\subsection{ADE 4--fold singularities}

Consider the polynomials
\bea
H = z_1^k + z_2^2 + z_3^2 & & A_{k-1}\nonumber\\
H = z_1^k + z_1z_2^2 + z_3^2 & & D_{k+1}\nonumber\\
H = z_1^3 + z_2^4 + z_3^2 & & E_6\nonumber\\
H = z_1^3 + z_1z_2^3 + z_3^2 & & E_7\nonumber\\
H = z_1^3 + z_2^5 + z_3^2 & & E_8~.\eea
The hypersurfaces $\{H=0\}\subset\C^3$ are known as the ADE 
surface singularities. Their links $L_{ADE}$ are precisely 
$S^3/\Gamma$ where $\Gamma\subset SU(2)$ are the finite ADE 
subgroups of $SU(2)$ acting on $\C^2$ in the vector representation. 
Thus these Gorenstein singularities are both 
hypersurface singularities and quotient singularities. 
Clearly, the links admit Sasaki--Einstein metrics -- they are 
just the quotient of the round metric on $S^3$ by the group $\Gamma$.

The 3--fold singularities of the previous section are obtained 
from the polynomial $H$ for $A_{k-1}$ by simply adding an additional 
term $z_4^2$ (and relabelling). More generally, we may define the ADE $n$--fold 
singularities as the zero loci $X=\{F=0\}$ of
\be
F = H+\sum_{i=4}^{n+1}z_i^2~.\ee
Let us consider the particular case $n=4$. The $\C^*$ actions, 
for the above cases, are generated by the weight vectors
\bea
\mathbf{w} = (2,k,k,k,k) \quad d=2k& & A_{k-1} \nonumber\\
\mathbf{w} = (2,k-1,k,k,k)\quad d=2k & & D_{k+1} \nonumber\\
\mathbf{w} = (4,3,6,6,6) \quad d=12 &  & E_6\nonumber\\
\mathbf{w} = (6,4,9,9,9) \quad d=18 & & E_7\nonumber\\
\mathbf{w} = (10,6,15,15,15) \quad d=30 & & E_8~.\eea
It is then straightforward to verify that for all the $A_{k-1}$ singularities 
with $k>3$ the holomorphic function $z_1$ on $X$ violates the Lichnerowicz bound (\ref{lichbound}). The case $k=3$ saturates the bound, but since the link is 
not\footnote{One can easily show that $H_3(L;\Z)=\Z_k$ for 
the links of the $A_{k-1}$ 4--fold singularities.} diffeomorphic to $S^7$ 
Obata's result \cite{obata} again rules this 
out. For all the exceptional singularities 
the holomorphic function $z_2$ on 
$X$ violates (\ref{lichbound}). The $D_{k+1}$ singularities 
are a little more involved. The holomorphic function $z_1$ 
rules out all $k>3$. On the other hand the function $z_2$ rules 
out $k=2$, but the Lichnerowicz bound is unable to rule out 
$k=3$. 

To summarise, the only ADE 4--fold singularity that might 
possibly admit a Ricci--flat K\"ahler cone metric with the canonical 
$\C^*$ action above, apart from the obvious cases of $A_0$ and $A_1$, 
is $D_4$. Existence of a Sasaki--Einstein metric on the link 
of this singularity is therefore left open. It 
would be interesting to investigate whether or
not there exist Ricci--flat K\"ahler metrics that are cones with
a \emph{different} Reeb action.
In light of our results on the non--existence of 
the above Sasaki--Einstein metrics,
it would also be interesting to revisit the field theory analysis of 
\cite{GVW}.

\subsection{Weighted actions on $\C^n$}

Consider $X=\C^n$, with a weighted $\C^*$ action with 
weights $\mathbf{v}\in (\Z_+)^n$. The orbit space of non--zero vectors is 
the weighted projective space $\mathbb{WCP}^{n-1}_{[v_1,\ldots,v_n]}$. 
Existence of a Ricci--flat K\"ahler cone metric on 
$\C^n$, with the conical symmetry generated by this $\C^*$ action, 
is equivalent to existence of a K\"ahler--Einstein orbifold 
metric on the weighted projective space. In fact, it is well known 
that no such metric exists: the Futaki invariant of the 
weighted projective space is non--zero. In fact, one can see this 
also from the Sasakian perspective through the results of \cite{MSY, MSY2}. 
The diagonal action with weights $\mathbf{v}=(1,1,\ldots,1)$ is clearly 
a critical point of the Sasakian--Einstein--Hilbert action, and this 
critical point was shown to be unique in the space of toric Sasakian 
metrics. Nonetheless, in this subsection we show that 
Lichnerowicz's bound and Bishop's bound 
both obstruct existence of these metrics. 

The holomorphic $(n,0)$--form on $X=\C^n$ has charge 
$|\mathbf{v}|$ under the weighted $\C^*$ action, which implies 
that the correctly normalised Reeb vector field is 
\be
\xi = \frac{n}{|\mathbf{v}|}\zeta
\ee
with notation as before, so that $\zeta$ is the vector field 
that generates $U(1)\subset\C^*$. The Lichnerowicz bound is therefore
\be
\frac{nv_m}{|\mathbf{v}|}\geq 1\ee
where $v_m$ is the (or a particular) smallest weight. However, 
clearly $|\mathbf{v}|\geq nv_m$, with equality if and only if 
$\mathbf{v}$ is proportional to $(1,1,\ldots,1)$. Thus in fact
\be
\frac{nv_m}{|\mathbf{v}|}\leq 1\ee
with equality only in the diagonal case, which is just $\C^n$ with the 
canonical Reeb vector field. Thus our Lichnerowicz bound obstructs 
K\"ahler--Einstein orbifold metrics on all weighted projective spaces,
apart from $\mathbb{CP}^{n-1}$ of course.

For the Bishop bound, notice that a K\"ahler--Einstein orbifold 
metric on $\mathbb{WCP}^{n-1}_{[v_1,\ldots,v_n]}$ would give rise to 
a Sasaki--Einstein metric on $S^{2n-1}$ with a weighted Reeb action. 
The volume of this metric, relative to the round sphere, would be
\be\label{volweighted}
V = \frac{|\mathbf{v}|^n}{n^nv}\ee
where $v=\prod_{i=1}^n v_i$ denotes the product of the weights. 
This may either be derived using the methods described earlier, 
or using the toric methods of \cite{MSY}.
Amusingly, (\ref{volweighted}) is precisely the arithmetic mean 
of the weights $\mathbf{v}$ divided by their geometric mean, all to 
the $n$th power. Thus the usual arithmetic mean--geometric mean 
inequality gives $V\geq 1$ with equality if and only if $\mathbf{v}$ 
is proportional to $(1,1,\ldots,1)$. This is precisely opposite 
to Bishop's bound, thus again ruling out all weighted projective spaces, 
apart from $\mathbb{CP}^{n-1}$.

Thus K\"ahler--Einstein orbifold metrics on 
weighted projective spaces are obstructed by the Futaki invariant, 
the Bishop obstruction, and the Lichnerowicz obstruction. 
In some sense, these Fano orbifolds couldn't have more wrong with them.

\section{Conclusions}
\label{concl}

The problem of existence of conical Ricci--flat K\"ahler 
metrics on a Gorenstein $n$--fold singularity $X$
is a subtle one; a set of necessary and sufficient algebraic 
conditions is unknown. 
This is to be contrasted with the case of \emph{compact} 
Calabi--Yau manifolds, where Yau's theorem 
guarantees the existence of a unique Ricci--flat K\"ahler metric 
in a given K\"ahler class.

In this paper we have presented two simple necessary conditions for 
existence of a Ricci--flat K\"ahler cone metric on a given 
isolated Gorenstein singularity $X$ with specified Reeb vector field. 
The latter is in many ways similar to specifying a ``K\"ahler class'', 
or polarisation. 
These necessary conditions
are based on the classical results of Bishop and  Lichnerowicz, that 
bound the volume and the smallest eigenvalue of the Laplacian on 
Einstien manifolds, 
respectively. The key point that allows us to use these as 
obstructions is that, in both cases, fixing a putative Reeb 
vector field $\xi$ for the Sasaki--Einstein metric is sufficient 
to determine both the volume and the ``holomorphic'' eigenvalues using 
only the holomorphic data of $X$. 
Note that any such vector field 
$\xi$ must also be a critical point of the Sasakian--Einstein--Hilbert
action of \cite{MSY,MSY2}, which in K\"ahler--Einstein terms means that the 
transverse Futaki invariant is zero. We emphasize, however, that the 
possible obstructions presented here may be analysed independently of 
this, the weighted projective spaces at the end of section \ref{section5} being examples 
that are obstructed by more than one obstruction.

To demonstrate the utility of these criteria, 
we have provided many explicit examples of Gorenstein singularities 
that \emph{do not} admit Sasaki--Einstein metrics on their links, for a particular 
choice of Reeb vector field. 
The examples include various quasi--homogeneous 
hypersurface singularities, previously studied in the physics literature, that 
have been erroneously assumed to admit such Ricci--flat K\"ahler cone metrics. 

We expect that in the particular case that the singularity is toric, 
neither Lichnerowicz nor Bishop's bound will obstruct for the 
critical Reeb vector field $\mathbf{b}_*$ of \cite{MSY}.
This is certainly true for all cases that have been analysed in the 
literature. In this case both bounds reduce
to simple geometrical statements on the polyhedral cone ${\cal C}^*$ and its associated semi--group
${\cal S}_{\cal C}={\cal C}^*\cap \Z^n$. For instance, given the critical 
Reeb vector field $\mathbf{b}_*$, 
the Lichnerowicz bound implies that $(\mathbf{b}_*,\mathbf{m})\geq 1$ for all 
$\mathbf{m} \in {\cal S}_{\cal C}$. It would be interesting to try to 
prove that this 
automatically follows from the extremal problem in \cite{MSY}, for
any toric Gorenstein singularity.

We have also explained the relevance of these bounds
to the AdS/CFT correspondence.
We have shown that the 
Lichnerowicz bound is equivalent to the unitarity 
bound on the scaling dimensions of BPS chiral operators of 
the dual field theories. In particular, we analysed a 
class of obstructed 
3--fold singularities, parameterised by a positive integer $k$,
for which, in the case that $k$ is even, the field theory dual is 
known and has been extensively studied in the literature. 
The fact that the links $L_k$ do not admit Sasaki--Einstein metrics 
for any $k>3$ supports the field theory arguments of \cite{Halmagyi}. 
It would be interesting to know whether a Sasaki--Einstein metric exists on
$L_3$; if it does exist, it might be dual to an exotic type of field theory 
since the corresponding Calabi--Yau cone does not admit a crepant resolution.
For the 4--folds studied in \cite{GVW}, it will be interesting to analyse the implications of our
results for the field theories.

\subsection*{Acknowledgments}
\noindent 
We would like to thank O. Mac Conamhna and especially D. Waldram for collaboration in the 
early stages of this work.
We would also like to thank G. Dall'Agata, 
M. Haskins, N. Hitchin, K. Intriligator, P. Li, R. Thomas, C. Vafa, 
N. Warner, and S. S.--T. Yau
for discussions. We particularly thank R. Thomas for comments 
on a draft version of this paper. J. F. S. is supported by NSF grants DMS--0244464, DMS--0074329 and DMS--9803347.
 S.--T. Y. is supported in part by NSF grants DMS--0306600 and DMS--0074329.
 
\appendix

\section{Cohomogeneity one metrics}

Here we discuss the equations that need to be solved to obtain 
a K\"ahler--Einstein orbifold metric on the Fano orbifold $V_k$ of 
section \ref{3fold}, which recall is a hypersurface 
$F=z_1^2+z_2^2+z_3^2+z_4^k=0$ in the weighted projective 
space $\mathbb{WCP}^3_{[k,k,k,2]}$. The group $SO(3)$ 
acts on $z_i$, $i=1,2,3$, in the vector representation, and 
then Matsushima's theorem \cite{Matsushima} implies that this 
acts isometrically on any K\"ahler--Einstein metric. The generic 
orbit is three--dimensional, and hence these metrics are 
cohomogeneity one. The K\"ahler--Einstein condition then reduces 
to a set of ordinary differential equations in a rather standard way. 

The ODEs for a local K\"ahler--Einstein 4--metric with cohomogeneity one 
$SU(2)$ action have been written down in \cite{dancer}. The metric may be 
written as
\be
\diff s^2 = \diff t^2 + a^2(t)\sigma_1^2+b^2(t)\sigma_2^2 
+ c^2(t)\sigma_3^2\ee
where $\sigma_i$, $i=1,2,3$, are (locally) left--invariant one--forms on 
$SU(2)$, and $t$ is a coordinate 
transverse to the principal orbit. The ODEs are then \cite{dancer}
\bea
\frac{\dot{a}}{a} &=& -\frac{1}{2abc}(b^2+c^2-a^2)\nonumber\\
\frac{\dot{b}}{b} &=& -\frac{1}{2abc}(a^2+c^2-b^2)\nonumber\\
\frac{\dot{c}}{c} &=& -\frac{1}{2abc}(a^2+b^2-c^2)+\Lambda 
\frac{ab}{c}
\label{odes}
\eea
where $\Lambda$ is the Einstein constant, which is $\Lambda=6$ in 
the normalisation relevant for Sasaki--Einstein metrics on $L_k$. 

The key question, given these local equations, is what the boundary 
conditions are. For a complete metric on $V_k$, the parameter $t$ must take 
values in a finite interval, which without loss of generality we may take 
to be $[0,t_*]$ for some $t_*$. At the endpoints, the principal orbit 
collapses smoothly (in an orbifold sense) to a special orbit. It is not difficult to work out 
the details for $V_k$, given its embedding in $\mathbb{WCP}^3_{[k,k,k,2]}$. 
One must separate $k=2p$ even and $k$ odd. 

For $k$ odd, the principal orbit is $SO(3)/\Z_2$. This collapses to 
the two special orbits
\bea
B_{t=0} & = & \left(SO(3)/\Z_2\right)/U(1)_1 =\mathbb{RP}^2 \nonumber \\
B_{t=t_*} & = & \left(SO(3)/\Z_2\right)/U(1)_3 = \mathbb{CP}^1\eea
where the circle subgroups $U(1)_1, U(1)_3\subset SO(3)$ are rotations 
about the planes transverse to the 
$1$--axis and the $3$--axis, respectively, thinking of $SO(3)$ acting 
on $\R^3$ in the usual way. Thus the two $U(1)$ 
subgroups are related by a conjugation. 

For $k=2p$ even, the principal orbit is instead simply $SO(3)=\mathbb{RP}^3$.
The two special orbits are 
\bea
B_{t=0} & = & SO(3)/U(1)_1 =S^2 \nonumber \\
B_{t=t_*} & = & SO(3)/U(1)_3 = \mathbb{CP}^1~.\eea
Of course, these are diffeomorphic, but the notation indicates that 
the second orbit is embedded as a complex curve in $V_{2p}$, whereas 
$S^2$ is embedded as a \emph{real} submanifold of $V_{2p}$.

In both bases, with $k$ odd or $k=2p$ even, 
the bolts are the real section of $V_k$, and the subvariety $z_4=0$, 
respectively. The latter is the image of the conic in $\mathbb{CP}^2\subset 
\mathbb{WCP}^3_{[k,k,k,2]}$ at $z_4=0$, and is a locus of orbifold
singularities. This is the only singular set on $V_k$.

The boundary conditions at $t=0$ are then, in all cases,
\bea
a(t) & = & \beta +\mathcal{O}(t)\nonumber \\
b(t) & = & \beta + \mathcal{O}(t)\nonumber \\
c(t) & =& \frac{2}{k} t +\mathcal{O}(t^2)\eea
where 
\be 
\beta^2 = \frac{k+2}{6k}~.\ee
At $t=t_*$, one simply requires that $a$ collapses to zero $a(t_*)=0$,  
with $b(t_*)=c(t_*)$ positive and finite. The metric functions should 
remain strictly positive on the open interval $(0,t_*)$.

The system of first order ODEs (\ref{odes}) may be reduced to a single second order
ODE as follows. The change of variables $\diff r/ \diff t= 1/c$ allows one 
to find the integral 
\be
\frac{a}{b}(r) = - \coth (r)
\ee
where an integration constant can be reabsorbed by a shift of $r$. Defining $f(r)=a b$, one obtains
\be
\frac{\diff}{\diff r} \log \left(f\frac{\diff f}{\diff r}\right) = 2 \left[\Lambda f +\coth(2r)\right]~.
\ee
Any solution of this equation 
gives rise to a solution of (\ref{odes}), using the fact that 
\be
c^2 = -\frac{\diff f}{\diff r}~.
\ee
For $k=1$, $k=2$, one can write down explicit solutions to these
equations and boundary conditions, corresponding to the 
standard metrics on $\mathbb{CP}^2$ and $\mathbb{CP}^1\times\mathbb{CP}^1$, 
respectively. 
For  $k=1$ we have
\be
a(t)=\cos \left(t+\frac{\pi}{4}\right),\qquad b(t)=\sin \left(t+\frac{\pi}{4}\right),\qquad
c(t)=\sin (2t)
\ee 
where the range of $t$ is $0\le t\le\pi/4$. Correspondingly,
\be
f(r) = -\frac{1}{2}\tanh (2r)
\ee 
with $\tan(t)=\exp{(2r)}$, so that $-\infty\leq r\leq 0$. 

For $k=2$ we instead have
\be
a(t)=\frac{1}{\sqrt{3}}\cos (\sqrt{3}t),\qquad b(t)=\frac{1}{\sqrt{3}},\qquad
c(t)=\frac{1}{\sqrt{3}}\sin (\sqrt{3}t)
\ee 
where the range of $t$ is $0\le t\le\pi/(2\sqrt{3})$. Correspondingly,
\be
f(r) = -\frac{1}{3}\tanh (r)
\ee 
with $\tan(\sqrt{3}t/2)=\exp{(r)}$ and $-\infty\leq r\leq 0$.

For all $k>3$, this paper implies that 
there do not exist any solutions. This still leaves the case $k=3$. 
We have neither been able to integrate the equations explicitly, 
nor have our preliminary numerical investigations been conclusive. 
We leave the issue of existence of this solution open.

\end{document}